# FLUXON-INDUCED LOSSES IN NIOBIUM THIN-FILM CAVITIES REVISITED*

W. Weingarten†

## Abstract

Long standing data from niobium thin film accelerating cavities will be revisited and analysed by a modified London model of RF superconductivity. Firstly, the applicability of this model is explored using data of the BCS surface resistance and its dependence on the RF magnetic field, temperature and mean free path. Secondly, the RF losses from trapped magnetic flux are analysed with regard to their dependence on these same parameters.

## INTRODUCTION

The two-fluid model of Gorter and Casimir [1] was extended by London for RF applications [2], many years before the Bardeen-Cooper-Schrieffer (BCS) theory of superconductivity [3] was published. The two-fluid model describes the Meissner effect [4] and also, though in a qualitative way and after some modification, the surface resistance $R_s$ of classical superconductors ("modified London model", cf. eq. 1).

The surface resistance $R_s$, or equivalently, the Q-value ($R_s \sim Q^{-1}$), are important parameters for accelerator application with respect to cryogenic losses and beam stability [5]. Therefore, in this paper, the modified London model will be applied to debate the RF field dependence of the Q-value on trapped magnetic flux of a 1.5 GHz thin niobium on a copper substrate cavity (with 1 - 2 μm thick niobium film) by using data from Benvenuti et al. [6]. The present analysis also constitutes a follow-up of, and a complement to, a previously published study [7].

The guiding principle of this paper consists in explaining available experimental data by minimum physics arguments as basic as reasonably possible. This is why, for instance, in the appendix the lumped electrical circuit model is used to derive the surface resistance due to fluxons.

The paper is organized as such: in the *first* section the modified London model is applied using data of the RF field dependent BCS surface resistance versus temperature and mean free path. In the *second* section the relation of the surface resistance on trapped magnetic flux will be analysed, both for the RF field independent and the RF field dependent part. The *third* section deals with the trapped magnetic flux induced surface resistance vs. temperature. The *fourth* section provides a side remark on fluxon sensitivity of N-doped cavities. A *fifth* section presents a critical review.

## THE RF-FIELD DEPENDENCE OF THE BCS SURFACE RESISTANCE ON TEMPERATURE AND MEAN FREE PATH

As a first test the data on the BCS-surface resistance $R_{BCS}$ will be analysed to gain confidence in the modified London model approach. We start with eq. 17 of ref. 8:

$$R_{BCS}(\omega,T) = \mu_0^2 \omega^2 \lambda^3 \sigma_n \overbrace{\frac{\Delta}{k_B} ln\left(\frac{\Delta}{\hbar\omega}\right) \frac{e^{-\frac{\Delta}{T}}}{T}}^{f'(T)} (1+ \alpha B_{rf} + \beta B_{rf}^2); B_c = 1/\sqrt{2\beta} \quad (1)$$

The linear term in $B_{rf}$ is suggested by ref. 6 and will be justified later in this paper, the quadratic term in $B_{rf}$ follows from experimental data [6, 9] and from different analytical models [8, 10].

Table 1: Fit parameters with regard to Fig. 1

| $\lambda$ (l) [nm] | $\sigma_n$ (RRR) [1/(Ωm)] | $\Delta$ [K] | $\alpha$ [1/(mT)] |
|---|---|---|---|
| 40 | 1.53·10⁸ | 18.9 | 7.5·10⁻³ |

The symbols are the peak RF magnetic field $B_{rf}$, the magnetic constant $\mu_0$, the frequency $\omega/(2\pi)$, the penetration depth $\lambda$, the electrical conductivity $\sigma_n$ of the normal conducting (nc) electrons at low temperature, their temperature dependence $f'(T)$ in the superconducting (sc) state, the Boltzmann constant $k_B$, and the sc energy gap $\Delta$. The (coloured) curves as shown in Fig. 1 follow from eq. 1 with the fit parameters as of Table 1. The critical magnetic field $B_c = 183$ mT (or $\beta = 1.5\cdot10^{-5}$ (mT)⁻², resp.) is kept fix.

Table 2: *Standard* parameters for niobium

| Parameter | Symbol | Value | Unit |
|---|---|---|---|
| Coherence length | $\xi_0$ | 38 | nm |
| London penetration depth | $\lambda_L$ | 39 | nm |
| Electrical conductivity at room temperature | $\sigma_n$ (300 K) | 7.6·10⁶ | (Ωm)⁻¹ |
| Electron mean free path | $l$ | 2.85·RRR | nm |
| Mass of electron | $m_e$ | 9.1·10³¹ | kg |
| Flux quantum | $\Phi_0$ | 2·10⁻¹⁵ | Vs |

---

The intrinsic parameters of niobium that are used for fitting the data of Fig. 1 are the London penetration depth $\lambda_L$ = 32 nm, the coherence length $\xi_0$ = 33 nm, and the electron mean free path $l$ = 57 nm.

In what follows, the *standard* parameters as of Table 2 are used unless otherwise stated.

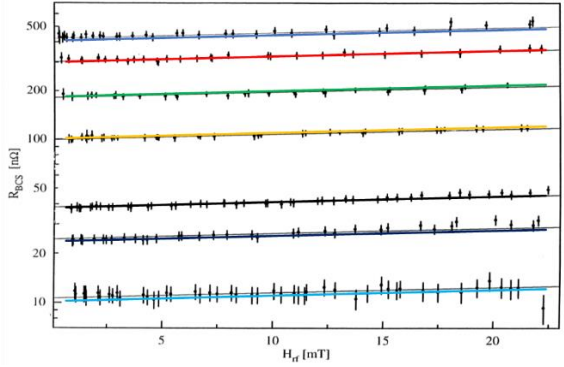

Fig. 1: The BCS-surface resistance $R_{BCS}$ of a thin niobium film cavity as a function of the magnetic peak RF field $H_{rf}$ for different temperatures (from top to bottom at 4.23, 3.9, 3.47, 3.07, 2.59, 2.41, and 2.15 K). Superimposed *in coloured lines* is a least square fit as suggested by eq. 1. The data are taken from ref. 6.

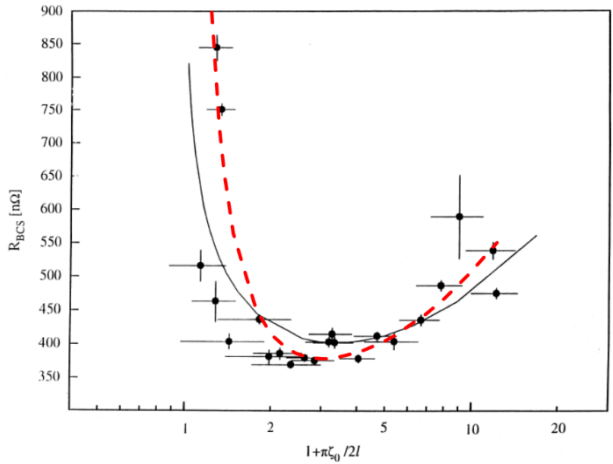

Fig. 2: BCS surface resistance vs. the relative penetration depth $(\lambda/\lambda_L)^2$ at 4.2 K. The dashed line (dashed red) is calculated from eq. 1 with slightly different $\lambda_L$ and $\xi_0$ as in ref. 6 and superimposed on the data from ref. 6 (all in black).

As a second test, the data on $R_{BCS}$ from ref. 6 are shown versus the square of the relative penetration depth: $\lambda_{rel}^2 = (\lambda/\lambda_L)^2 = 1 + \pi \cdot \xi_0/(2 \cdot l)$, Fig. 2. Superimposed is the result as derived from eq. 1. Here the relevant parameters are the same as for Fig. 1. The characteristic minimum is clearly visible at $\lambda_{rel}^2 = 3$ which corresponds to $l$ = 26 nm.

# THE DEPENDENCE OF THE SURFACE RESISTANCE ON TRAPPED MAGNETIC FLUX

The niobium thin film cavities developed at CERN are less sensitive to DC trapped magnetic flux $B$ when cooled down compared to cavities made from bulk niobium. The small dependence of the magnetically induced surface resistance $R_{fl}$ on $B$ and $B_{rf}$ can be parametrized as [6]

$$R_{fl} = (R_{fl}^0 + R_{fl}^1 \cdot B_{rf}) \cdot B \quad , \quad (2)$$

which is composed of the RF-field independent and the RF-field dependent fluxon sensitivities $R_{fl}^0$ and $R_{fl}^1$, resp., measured in nΩ/Gauss and nΩ/Gauss/mT, resp.

The losses from $R_{fl}^0$ may be understood by the voltage created from the inertia of the sc shielding current density $j$ which develops across the nc core of the trapped fluxons, as derived in ref. 7 (c.f. appendix). The current flows via two parallel impedances, one a resistance, the other an inductance. Nonetheless, ref. 7 merits to be revisited, because the postulated data for the upper critical field $B_{c2}$ of niobium are debateable and the RF magnetic field dependent contribution to the surface resistance $R_{fl}^1$ is not treated.

## The RF field independent contribution $R_{fl}^0$

In all what follows, the fluxons are considered to move freely, their depinning frequency being smaller than the RF frequency (1.5 GHz) [11]. This conjecture can be called into doubt because of two reasons. First, for small mean free paths ($l$ < 30 nm) the depinning frequency may well be near and above the GHz region [12]. Second, the depinning frequency may well depend on the specific pinning potential [13]. Nevertheless, as the surface defects are known as the main sources of pinning [14] and hence not uniformly distributed, the following study deals with a rigid lattice of pinning sites as originally proposed by Gittleman and Rosenblum [11]. The pinning potential then depends on the distance of the fluxons which on its part is based on the trapped magnetic field.

Whether this supposition is justified, was checked by evaluating the depinning frequency $f_d$ from eq. 3, adopted from ref. 11,

$$f_d = \frac{\rho_n \cdot \alpha}{\sqrt{\Phi_0 \cdot B \cdot B_{c2}}} \quad , \quad (3)$$

with $\rho_n$ being the electrical resistivity at low temperature, $\rho_n = (\sigma_{n,300K} \cdot RRR)^{-1}$, $\alpha$ the maximum Lorentz force the fluxon lattice can withstand, $\alpha = B \cdot J_c$, and $B_{c2}$ the upper critical magnetic field of the niobium film. For a quantitative analysis the following numbers are used, apart from the *standard* ones: $B$ = 50 (mGauss), about 10 percent of the earth magnetic field; $J_c \approx 2.5 \cdot 10^{10}$ (A/m$^2$)·(1 - $B$[T]/0.4) [15];-and $B_{c2}$ as depicted in Fig. 3.

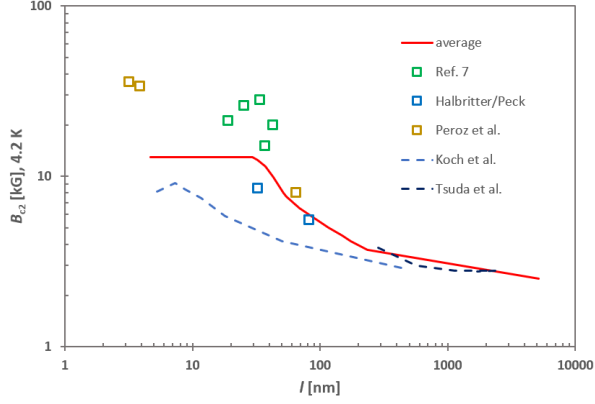

Fig. 3: Upper critical magnetic field $B_{c2}$ of samples of niobium thin films (*squares*) and bulk (*dashed lines*) vs. mean free path (derived from their *RRR* value). The red *solid* line represents "averaged" thin film data. The data for $l < 10$ nm were not taken into account because of their critical temperature $T_c \leq 8.9$ K much smaller than the observed one, 9.13 K $\leq T_c \leq$ 9.56 K [6].

The depinning frequency $f_d$ vs. the mean free path $l$ is shown in Fig. 4, top. The trapped-flux surface resistance $R_s$ is normalized to the one of the freely moving fluxon lattice and is shown vs. the RF frequency in Fig. 4, bottom. $R_s$ follows the relation $R_s/R_s(f \gg f_d) = f^2/(f^2 + f_d^2)$ [11]. Hence the supposition of freely moving fluxons is justified for the RF frequency of 1.5 GHz, even in the extreme of a very short mean free path of 5 nm (corresponding to $f_d = 77$ MHz). This statement is in accordance with studies of Janjušević and co-authors [16], who find that 160-nm niobium films have a depinning frequency below 1 GHz (with $RRR = 40$), which is supposed to be true as well for thicker films (with $7 < RRR < 29$ [17]), such as the 1- 2 μm thick niobium films as investigated in this study.

However, according to ref. 12, the depinning frequency is by more than one to two orders of magnitude larger than depicted in Fig. 4, top. The reason is the fixed and deep pinning potential as used in their study in contrast to the rigid lattice approach chosen in the present study. A sharper criterion cannot be given at present except whether the model is capable of representing the data or not.

There are two contributions to the RF field independent surface resistance $R_{fl}^0$. The **first contribution** is attributed to fluxons directly exposed to the RF shielding current density $j = (j_x, 0, 0)$.

As outlined in the appendix,

$$R_{fl}^0 = c \cdot (\omega\mu_0)^{3/2}(2\sigma_n)^{1/2}\lambda^2 \frac{1}{B_{c2}} \qquad . \quad (4)$$

The correction factor $c$ (62.5 %) takes into account the ratio of the magnetic flux component perpendicular to the cavity surface with regard to the overall magnetic flux across the cavity silhouette.

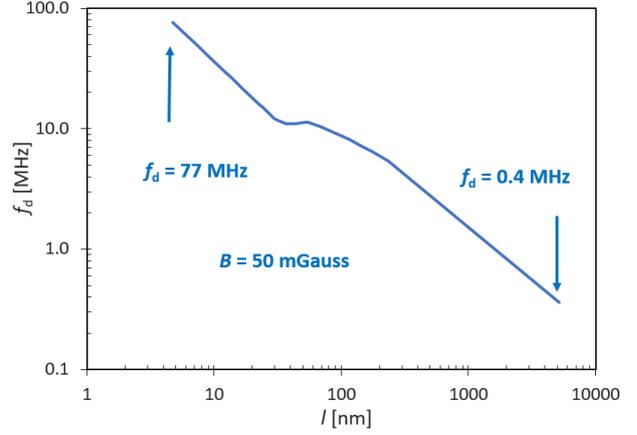

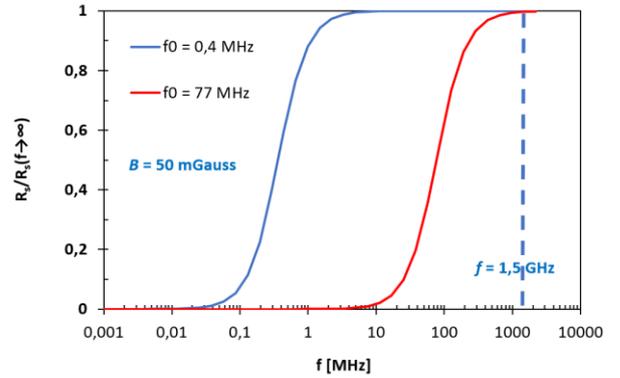

Fig. 4: Depinning frequency $f_d$ vs. mean free path $l$ (top), surface resistance $R_s$ from trapped flux, normalized to the one of the freely moving fluxon lattice (when $f \gg f_d$) vs. the RF frequency $f$ (bottom).

The **second contribution** is attributed to fluxons (indirectly) exposed to an inductive current. It is well known that an RF current density $j = (j_x, 0, 0)$ flowing perpendicular to a *static magnetic* field $B = (0, B_y, 0)$ will create the Lorentz force density on the fluxon $F = (0, 0, F_z) = j \times B$ that will move it with the velocity $v = \eta^{-1} \cdot F$ (c.f. Fig. 5 and Table 3).

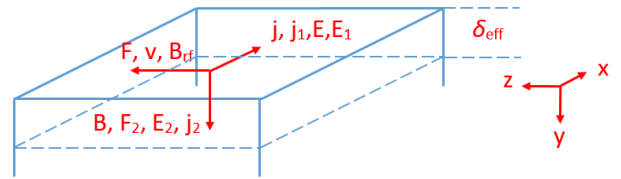

Fig. 5: Geometry as referred to in the text (*the letters indicate to which axis the corresponding vectors are parallel; they do not indicate their direction*)

The moving fluxon induces an electric field $E = (E_x, 0, 0) = B \times v$ that will create a current density $j_1 = (j_{1x}, 0, 0) = (E_{1x}/\rho_n, 0, 0)$.

The current $j_1$ acts on the nc electrons in a similar way as the current $j$, but in quadrature. Hence $j_1$ contributes identically to the fluxon induced surface resistance $R_{fl}^0$. The reason is that the force $F$ is in phase with $j$, as is the velocity $v$. But by induction, $v$ induces an electric field $E_1$ in quadrature to $j$. As consequence from Fig. 5, the different power dissipations $P$ and $P_1$ due to $j$ and $j_1$ (or equivalently $I$ and $I_1$ respectively), may be added: $P = \frac{1}{2} R (I + i \cdot I_1) \cdot (I - i \cdot I_1) = \frac{1}{2} R I^2 + \frac{1}{2} R I_1^2 + \frac{1}{2} i \cdot R \cdot I \cdot I_1 - \frac{1}{2} i \cdot R \cdot I \cdot I_1 = P+P_1$. This is also true for the respective surface resistances $R_s = E/(\lambda \cdot j) = E_1/(\lambda \cdot j_1)$. Hence $R_{fl}^0$ is composed of twice the value of eq. 4.

Table 3: Used symbols and their definition [18]

| Physical quantity | Symbol | Unit |
|---|---|---|
| Shielding current density | $j = E/\rho_n$ [1] | A/m$^2$ |
| Magnetic induction | $B$ | Vs/m$^2$ |
| Lorentz force density | $F = j \times B$ | N/m$^3$ |
| Fluxon velocity | $v = \eta^{-1} F$ [2] | m/s |
| Electric field from moving fluxons | $E = B \times v = \eta^{-1} B \times (j \times B) = \rho_{ff} j$ [3] | V/m |
| Electric field from Lorentz force density | $E = j \times B/(n \cdot e)$ [4] | V/m |
| Hall resistivity | $\rho_{yx} = E_y/j_x = R \cdot B$ [5] | Ωm |

[1] $\rho_n$ is the normal state resistivity at low temperature; [2] $\eta$ is the drag coefficient; [3] $\rho_{ff} \approx (B/B_{c2}) \rho_n$ [19];
[4] $n$ is the normal state electron density; [5] $R = 1/(n \cdot e)$ is the Hall coefficient

The fluxon sensitivity $R_{fl}^0$ is equivalent to the DC result for the "ideal" material as outlined by Gittleman and Rosenblum [11]. However, their fluxon sensitivity $R_s'$ is too large as to represent the data of ref. 6.

In order to make use of eq. 4, data of $B_{c2}$ for representative thin films similar to those grown on the cavity surface are collected from the literature (Table 4).

These data are plotted in Fig. 3 in conjunction with data on bulk niobium samples [20, 21] (dashed lines). The by-eye-averaged line of thin film data is used in the following analyses (marked as "average"). Whereas the data between $l = 3$ and 4 nm are disregarded because of a too small film thickness, a constant value is assumed below $l = 30$ nm as average of the data from refs. 7, 20, and 24.

Applying the average data (Fig. 3) to eq. 4 results in the red dashed curve of the trapped fluxon sensitivity $R_{fl}^0$ as of Fig. 6. The relevant parameters are those mentioned underneath Table 1.

The agreement with the published data of ref. 6 is satisfactory except for $l < 17$ nm ($\lambda_{rel}^2 > 4$), very probably because of the uncertain $B_{c2}$ in this range. The trend of the curve is well represented. It should be noted that this curve was obtained by taking into account the variation of the trapped flux density across the cavity surface. For the static magnetic field parallel to the cavity axis, the average flux density is 62.5 % as compared to a fictitious maximum flux density when all surface were exposed to the *perpendicular* component of the static magnetic field [22]. This correction shifts the curve slightly down and will be applied in what follows, too. Principally unknown is the trapping efficiency, but from experiment it is known to be close to one [23].

Table 4: Data of $B_{c2}$ for niobium thin films

| Film thickness [μm] | $B_{c2}$, 4.2 K [kGauss] | RRR | Mean free path $l$ [nm] = 2.85 ·RRR | Reference |
|---|---|---|---|---|
| 5 | 20 | 15 | 43 | adopted from ref. 7 |
| 3.7 | 21 | 6.7 | 19 | |
| 3 | 15 | 13 | 37 | |
| 3 | 26 | 9 | 26 | |
| 1.6 – 1.8 | 28 | 12 | 34 | |
| 1.5 | 8.5 | 11.5 | 33 | [24] |
| 1.5 | 5.5 | 29 | 83 | |
| 0.1 | 36*) | 1.1 | 3.2 | [25] |
| 0.1 | 8*) | 22.8 | 65 | |
| 0.1 | 34*) | 1.4 | 3.9 | |

*) Numbers were extrapolated to 4.2 K

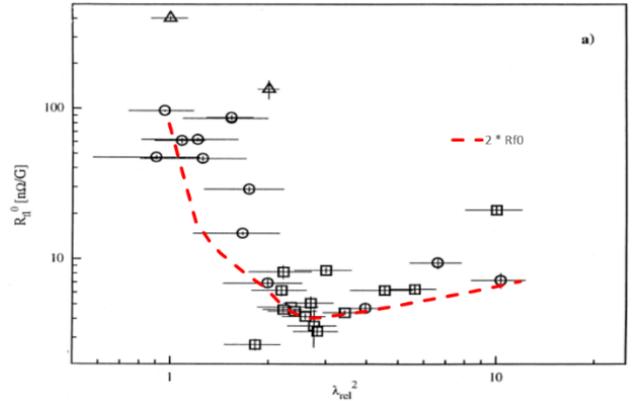

Fig. 6: Trapped fluxon sensitivity $R_{fl}^0$ vs. the square of the relative penetration depth $(\lambda/\lambda_L)^2$. The dashed line (red) is calculated from eq. 4 (but multiplied by factor 2, as explained in the text) using the "averaged" data for $B_{c2}$ and is superimposed on the data from ref. 6.

A different check of the model is provided by the dependence of the fluxon sensitivity $R_{fl}^0$ on the RF frequency $f = \omega/2\pi$. The data are obtained from Calatroni and Vaglio [26] and reproduced and supplemented in Table 5 and in Fig. 7.

The first four lines are measured for bulk niobium, while the two bottom lines for niobium films on copper.

Table 5: Magnetic flux sensitivities $R_{fl}^0$ and $R_{fl}^1$ as measured by several authors.

| Reference | Frequency [MHz] | $R_{fl}^0$ [nΩ/G] | $R_{fl}^1$ [nΩ/G/mT] |
|---|---|---|---|
| Piosczyk [27] | 91/160/290 | 3.5/9.5/28 | 0.35/0.55/0.9 |
| Arnolds-Mayer [28] | 500 | 150 | 5 |
| Checchin [12] | 650/1300/2600/3900 | 700/1000/1500/1900[1] | 1.6/2.6/6.1/7.4 |
| Benvenuti [29] | 1500 | 155 | 2.8 |
| Miyasaki [30] | 101 | 3.2 | 0.32 |
| Benvenuti [29] | 1500 | 3.3[2]/56 | 0.91[2]/4.5 |

[1]) Data taken from the original paper ref. 12
[2]) Thin niobium film on oxidized copper

Although the data on $R_{fl}^0$ were collected for a variety of experimental conditions, niobium metals, processing techniques, and in different laboratories, etc., they are not in contradiction with the expected frequency dependence ($\sim \omega^{3/2}$) of $R_{fl}^0$, c.f. eq. 4.

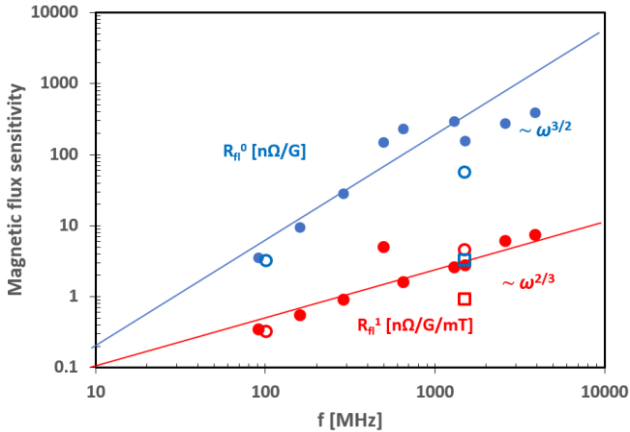

Fig. 7: Frequency dependence of the fluxon sensitivities $R_{fl}^0$ and $R_{fl}^1$ (full dots: bulk niobium; open dots: niobium film). Their dependence on the frequency suggest $\omega^{3/2}$ and $\omega^{2/3}$, resp., which is discussed in the text. The open squares fall out of the data collection and represent niobium film on oxidised copper cavities, known to have lower values $R_{fl}^0$ and $R_{fl}^1$ [29].

## *The RF field dependent contribution $R_{fl}^1$*

The RF field dependent part of the fluxon sensitivity $R_{fl}^1$ is actually under study by different authors [26, 30]. This part is observed in niobium bulk cavities [31] as well as in those that have undergone N-doping treatment [32].

### The role of the anomalous skin effect

The skin effect is created by surface currents in the metal which short-circuit the electric field parallel to the surface.

The domain of the anomalous skin effect is situated at sufficiently low temperatures and sufficiently high frequencies, where the mean free path $l$ of the electrons gets larger than the penetration depth [33]. Only electrons whose mean free path $l$ ranges within a surface layer where a non-vanishing electric field is present (the effective penetration depth $\delta_{eff}$) contribute to the current which shields the external RF field. The others are "invisible" to the electric field (Fig. 8) [34].

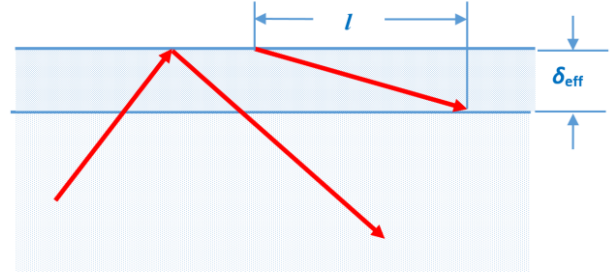

Fig. 8: Only electrons within the effective penetration depth contribute to shielding the external RF field.

Hence the effective density of the electrons is reduced by the factor $\alpha \cdot \delta_{eff}/l$ with $\alpha \approx 1$. The effective conductivity is, therefore, given by $\sigma_{eff} = \alpha \cdot \delta_{eff}/l \cdot \sigma_n$. Introducing this into the formula for the skin depth, $\delta = \sqrt{[2/(\mu_0 \cdot \sigma_n \cdot \omega)]}$, one obtains

$$\delta_{eff} = \left(\frac{2 \cdot l}{\alpha \cdot \mu_0 \cdot \sigma_n \cdot \omega}\right)^{1/3} \quad , \quad (5)$$

$$\sigma_{eff} = \left(\frac{2}{\mu_0 \cdot \omega}\right)^{1/3} \left(\frac{\alpha \cdot \sigma_n}{l}\right)^{2/3} \quad . \quad (6)$$

Similarly, the effective surface resistance is $R_{s,eff} = 1/(\sigma_{eff} \cdot \delta_{eff})$, which exhibits the characteristic frequency dependence ($\sim \omega^{2/3}$) of the surface resistance in the anomalous limit. The observed frequency dependence of $R_{fl}^1$ ($\sim \omega^{2/3}$) as of Fig. 7 already points to the anomalous skin effect as being relevant for $R_{fl}^1$.

Anyhow, it appears strange that the anomalous skin effect may play a role in the present analysis. However, the conditions for the anomalous skin effect are considered as being satisfied: the mean free path is large ($l > \lambda$) and the current carrying layer is very small ($< \xi$) as given by the fluxon width.

### The role of the Hall effect

It is observed that there exists a different electric field, the Hall field $\boldsymbol{E}_2 = (0, E_{2y}, 0)$, cf. Fig. 5, created by the Lorentz force density $\boldsymbol{F}_2 = (0, F_{2y}, 0) = \boldsymbol{j} \times \boldsymbol{B}_{rf} = (j_x, 0, 0) \times (0, 0, B_{rf,z})$ and concentrated in the vicinity of the fluxon. The electron feels the force $e\boldsymbol{E}_2 = \boldsymbol{F}_2/n = \boldsymbol{j} \times \boldsymbol{B}_{rf}/n$, $n$ being the electron density. This force creates the current $\boldsymbol{j}_2 = (0, j_{2y}, 0)$ with $j_{2y} = \sigma_{yx} \cdot E_x = E_x/(B_{rf,z} \cdot R) = n \cdot e/B_{rf,z} \cdot \rho_n \cdot j_x$, $R = 1/(ne)$ being the Hall coefficient (cf. Table 3). The RF losses per volume are then $P = \langle \boldsymbol{E}_2 \cdot \boldsymbol{j}_2 \rangle = \frac{1}{2} \cdot \rho_n \cdot j_x^2$. With $j_x = H_{rf}/\lambda$ there follows for the power loss per square meter $p$:

$$p = P \cdot \lambda = \frac{1}{2} \cdot \frac{\rho_n}{\lambda} H_{rf}^2 \quad .$$

With the surface fraction of fluxons $B/B_{c2}$ (cf. appendix), one obtains

$$p = \frac{1}{2} \cdot \frac{\rho_n}{\lambda} H_{rf}^2 \cdot \frac{B}{B_{c2}} = \frac{1}{2} \cdot R_{fl}^1 \cdot H_{rf}^2 \cdot B_{rf} \cdot B \quad,$$

resulting in

$$R_{fl}^1 = \frac{\rho_n}{\lambda \cdot B_{rf} \cdot B_{c2}} \quad . \quad (7)$$

Eq. 7 is now evaluated under similar parameters as for eq. 4, however with two distinctions. The first distinction is governed by the anomalous skin effect with the mean free path $l > \lambda$ (small $\lambda_{rel}$). The reason is that the driving electric field is localized near the moving fluxon. Therefore the electrons are exposed only partially to the electric field, a condition similar to the anomalous skin effect.

Hence the replacements $\rho_n \rightarrow 1/\sigma_{eff}$ and $\lambda \rightarrow \delta_{eff}$ from eqs. 5 and 6 are inserted in eq. 7.

For the second distinction ($l < \lambda$, big $\lambda_{rel}$, therefore $\rho_n = 1/\sigma_n$ is used in eq. 7) follows as such:

$$R_{fl}^1 = \frac{\rho_n}{\lambda \cdot B_{rf} \cdot B_{c2}} = \frac{1}{\sigma_n \cdot \lambda \cdot B_{rf} \cdot B_{c2}} = \frac{m}{n \cdot e^2 \cdot \tau \cdot \lambda \cdot B_{rf} \cdot B_{c2}} = \frac{1}{n \cdot e \cdot \underbrace{\tau \cdot \lambda \cdot \omega_c}_{l} \cdot B_{c2}} = \frac{1}{n \cdot e \cdot l \cdot B_{c2}} \quad (8)$$

(cyclotron frequency $\omega_c = e \cdot B_{rf}/m$, electrical conductivity $\sigma_n = ne^2\tau/m$, effective electron mass $m$, electron density $n$, electric charge $e$, collision time $\tau$). The mean free path $l \approx \tau \cdot \lambda \cdot \omega_c$ represents the typical length the electron can go without being scattered.

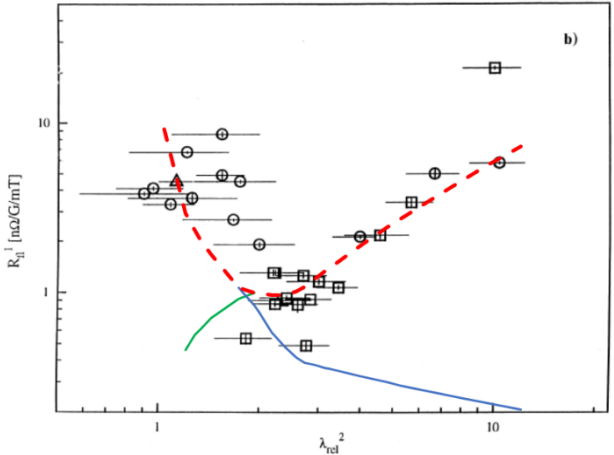

Fig. 9: Trapped fluxon sensitivity $R_{fl}^1$ vs. the square of the relative penetration depth $(\lambda/\lambda_L)^2$. The dashed line (red) is the combination of the two continuous lines (blue and green), as calculated from eqs. 7 and 8, and superimposed on the data from ref. 6.

Disordered niobium [35] or niobium alloys are considered to establish the fluxon pinning centres. Under this assumption 17 % of the *standard* value of $\sigma_n$ is kept fixed to fit the data for both distinctions. This provision indicates that the pinning centres are depleted from electrons. The result of the fit is shown in Fig. 9. There, the two distinctions are marked as a dashed line (left: $l > \lambda$; right: $l < \lambda$). The left branch with $l > \lambda$ shows the characteristic frequency dependence of $R_{fl}^1$, as characteristic for the anomalous skin effect for bulk niobium ($\sim \omega^{2/3}$), in accordance with Fig. 7.

### Combination of $R_{fl}^0$ and $R_{fl}^1$

As a check of the results obtained so far, Fig. 10, adopted from ref. 6, displays the combined fluxon sensitivity of eq. 2. The red line is superimposed by fitting these data with the parameters $R_{fl}^0 = 4.7$ nΩ/G and $R_{fl}^1 = 1.0$ nΩ/G/mT. These numbers are consistent with Figs. 6 and 9.

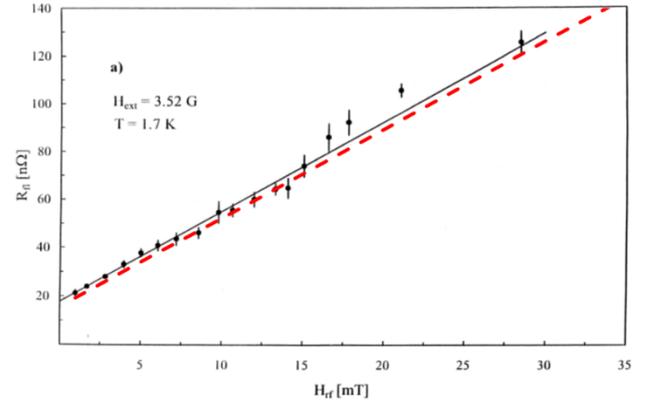

Fig. 10: Combined fluxon sensitivity vs. the RF magnetic field at an external magnetic field 3.52 Gauss (adopted from ref. 6).

The computed data from eqs. 4, 7 and 8 (Figs. 6 and 9, dashed lines) are correlated as shown in Fig. 11.

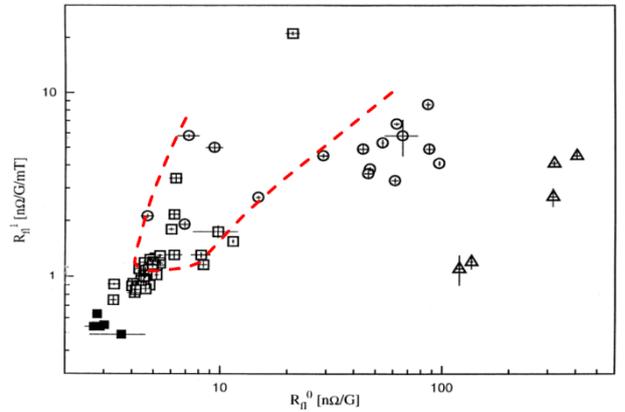

Fig. 11: Correlation of $R_{fl}^1$ vs. $R_{fl}^0$ (data adopted from ref. 6). The data points for $R_{fl}^0 > 100$ nΩ/G represent bulk cavities and are not representative.

## THE DEPENDENCE OF THE TRAPPED MAGNETIC FLUX INDUCED SURFACE RESISTANCE ON THE TEMPERATURE

Eqs. 4, 7 and 8 show that the fluxon sensitivity depends on $\lambda(T)$ and $B_{c2}(T)$. For the penetration depth $\lambda$ the standard relation is used,

$$\lambda = \lambda_L \cdot \frac{\sqrt{1+\frac{\pi \xi_0}{2l}}}{\sqrt{1-\left(\frac{T}{T'_c}\right)^4}} \qquad , \qquad (9)$$

with the free parameter $T_c$' allowing for disordered niobium or niobium alloys as mentioned before. For $B_{c2}(T)$ the relation $B_{c2}$ [T] = 4·(1-$T$ [K]/9.25) as presented in Fig. 4 of ref. 36 is applied, though with the caveat that $B_{c2}(T)$ for the present data is unknown.

The data to be analysed are shown in Fig. 12, which displays the ratio $r = R_{fl}(T)/R_{fl}$ (1.7 K), as defined in ref. 6. According to ref. 6 this graph is quite universal and independent of the specific choice of $B_{rf}$ and $B$. The ratio $r$ was measured at identical values of $B_{rf}$ and $B$ and then displayed as a function of the temperature. The dashed line in Fig. 12 was computed by means of eqs. 4 and 7 with the usual parameters as of the graph in Fig. 3 with $RRR$ = 20 and $B_{rf}$ = 5 mT and with $T_c$' = 4.65 K. The critical temperature $T_c$' is untypical for ordinary niobium and points indeed to disordered or contaminated niobium.

The analysis allows concluding that Fig. 12 reflects mainly the relatively strong dependence on the temperature $T$ of $R_{fl}^0$ because the penetration depth $\lambda$ as of eq. 9 is supposed to increase steeply above about $T_c$' ≈ 4.5 K. The dependence on $T$ of $R_{fl}^1$, on the contrary, is weak up to 4.2 K consequent to the relatively weak dependence of $B_{c2}$ on $T$.

## ACCESSORY REMARK

Other experimental results are worth debating in the context of this paper in order to illustrate the lumped circuit model approach. These experiments concern the effect of nitrogen doping on the trapped magnetic flux sensitivity of niobium cavities [31, 37, 38]. Instead of identifying a minimum for the fluxon sensitivity $R_{fl}^0$ vs. the mean free path (Fig. 6), these authors observe a maximum such as shown in Fig. 13. A detailed analysis of this apparent contradiction lies beyond the scope of the paper. Although the peak in the fluxon sensitivity is explained by well-established flux-pinning models [39, 40], a short and simple comment referring to this shall be given in what follows and in the appendix.

Applying eq. A-10 and minimizing the mean square deviation between data and fit one obtains typically a dashed red dashed curve as in Fig. 13. In this specific case the fit parameters are collected in Table 6.

Table 6: Numerical values concerning Fig. 13.

| Name | Value[4] | Unit | Ratio |
|---|---|---|---|
| $B$ [1] | 5·10$^{-5}$ | Vs/m$^2$ | - |
| $\sigma_n$ | (6.2±1.3)·10$^6$ | 1/$\Omega$m | 0.8 |
| $J_c$ [2] | 2.4·10$^{10}$ | A/m$^2$ | 10 |
| $n$ | (7.1±0.6)·10$^{27}$ | m$^{-3}$ | 0.13 |
| $B_{c2}(0)$ [3] | 8400±600 | Gauss | 2 |
| $\lambda_L$ | 87±14 | nm | 2 |
| $\xi_0$ | 59±5 | nm | 1.6 |

[1]) average earth magnetic field
[2]) *Standard* critical current density for Nb: $J_c$ = 2.5·10$^9$ (A/m$^2$) [41]
[3]) $B_{c2}(\rho_n) = B_{c2}(0) + m \cdot \rho_n$; with $B_{c2}(0)$ = 4040 Gauss [41] and $m$ =1.5·10$^3$ [Gauss/μΩcm] as average number [42, 43]
[4]) Errors added after proof, except for $J_c$, for which the error is unreasonably large

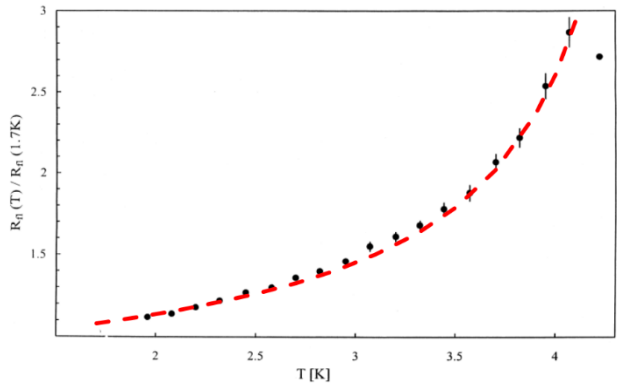

Fig. 12: Increase of the ratio $r$ of the fluxon sensitivity vs. temperature. The dashed line (red) is calculated from eqs. 3 and 7 and superimposed on the data from ref. 6.

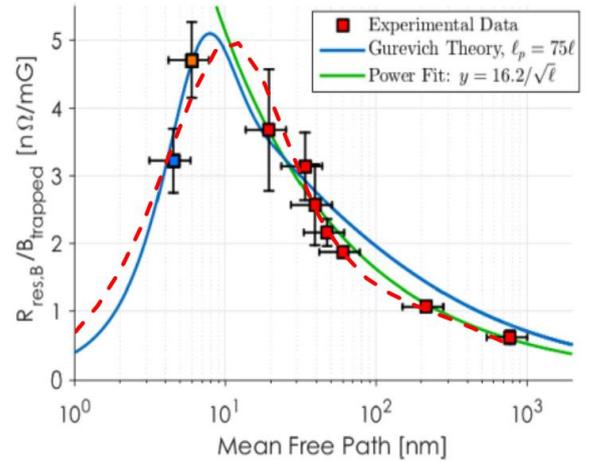

Fig. 13: Trapped fluxon sensitivity $R_{fl}^0$ for niobium cavities at 1.3 GHz (adopted from ref. 37). The superimposed dashed line (in red) results from eq. A-10 by using the data set of Table 6.

Inspecting Table 6, the last column provides the ratio between the actual fit parameter and its actual *standard* value. One may conclude that N-doped niobium has a smaller than *standard* electron concentration $n$ (as already met for thin film cavities) and larger critical current density $J_c$ as the most striking differences compared to *standard* niobium.

The shape of the graph of Fig. 13 resembles much that of ref. 38, Fig. 2 therein. The authors interpret the region left of the maximum as "pinning regime" and that right of it as "flux flow regime".

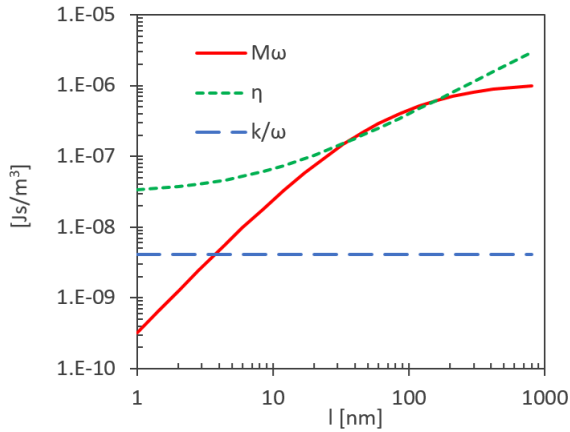

Fig. 14: Comparison of the three force terms (per unit fluxon length and unit velocity) as of eq. A-5 and related to Fig. 13; the RF frequency is 1.3 GHz.

Fig. 14 shows the juxtaposition of the various relevant force terms concordant with the numbers in Table 6. The mass term is irrelevant for a mean free path below a few nanometres, where the viscous ($\eta$) and pinning terms ($k/\omega$) are dominant. However, above a mean free path of about 10 nm the mass ($M \cdot \omega$) and viscous terms are the dominant contributors.

The depinning frequency is usually defined for frequencies below 1 GHz, which is fully justified because of the small mass term. But the notion of depinning as originally proposed in ref. 11 makes it difficult to explain a resonance peak as in Fig. 13. Therefore, in the appendix, the resonant peak is proposed to originate from oscillating fluxons near the resonant frequency $\omega_0 = \sqrt{k/M}$, close to where the pinning and mass terms intersect in Fig. 14. The idea of oscillating fluxons was already suggested before [39] in an analysis supposing strong pinning. The idea was not followed up in view of the large resonant frequency usually not applied for accelerator application.

## REVIEW

Admittedly the preceding analysis is based on a few deficiencies, such as the scarce knowledge of $B_{c2}$ for thin films. In addition, the actual trapping efficiency of the magnetic flux is unknown though considered as fairly complete from similar other experiments. In spite of these fragmentary information, the most interesting outcome from this (lumped circuit based) analysis is the fact that the observed RF losses, for niobium thin film cavities, can be best described by

- fluxons with local critical temperature around $T_c' \approx 4.5$ K and a reduced electron density (~ 17 %), compared to *standard* niobium,
- localized RF losses originating inside and in close vicinity of these fluxons,
- created by the moving fluxons and the local Hall field directed perpendicular to the current carrying surface, and
- the anomalous skin effect (for mean free paths larger than the penetration depth) due to the ineffectiveness of the shielding current along the fluxons.

That the RF losses are concentrated around, and dominated by, the fluxons is not surprising, because the surface resistance from different cavity wall areas is additive and hence naturally dominated by the lossiest areas. The associated local critical temperature $T_c'$ may hint on dirty and/or disordered niobium rich with dislocations, or on dissolved oxygen near the solubility limit. It is evident that the external static magnetic field will preferentially be trapped precisely there.

As to N-doped cavities, the lumped circuit based approach provides the following result. The observed peak in the fluxon sensitivity $R_{fl}^0$ may be caused by a resonant absorption from the exchange of current and charge inside the fluxons probably trapped near disorderd niobium.

## CONCLUSION

In this paper a modified London model of RF superconductivity allows quantifying the RF losses in sc niobium thin film cavities originating from trapped fluxons, considered as being depinned and hence mobile at the RF frequency under study (1.5 GHz).

The RF losses from trapped fluxons consist of two contributions, those *directly* exposed to the RF shielding current and those *indirectly* exposed to the RF inductive current. The directly exposed fluxons experience RF losses similar to nc defects across the current path. The indirectly exposed fluxons contribute to the RF losses in two ways. Firstly, they create a current in quadrature but parallel to the shielding current and hence give rise to the same additive surface resistance as the latter. Secondly, they create an RF Hall current perpendicular to the surface and confined within the small penetration depth, also dissipating energy in the fluxons. A model in accordance with these explanations corroborates the experimental facts of ref. 6: the surface resistance for both species of current increases linearly with the fluxon density, and that due to the Hall current increases linearly with the RF field amplitude. The minimum surface resistance from trapped fluxons is associated with *RRR* about 9 to 18.

# ACKNOWLEDGEMENT

The author thanks Sergio Calatroni (CERN) for valuable hints and suggestions.

# APPENDIX

### $R_{fl}^0$ for conventional (non N-treated) niobium cavities

The lumped-circuit model of Fig. 15 shows an inductance $L$ and a resistance $R$ in parallel subject to the total current $I$. The inductance describes the sc electrons, the inertia of which give rise to a voltage $V$. The resistance describes the nc ones present in the fluxon, subject to the voltage $V$.

Applying Kirchhoff's node rule at point "A",

$$I = I_1 + I_2 = V\left(\frac{1}{R} + \frac{1}{i\omega L}\right) \quad , \quad \text{(A-1)}$$

the complex impedance from the fluxon becomes

$$Z = \frac{V}{I} = \frac{1}{\left(\frac{1}{R} - \frac{i}{\omega L}\right)} \quad ,$$

the real part of which is

$$Re(Z) = \frac{R}{1 + \left(\frac{R}{\omega L}\right)^2} \quad ,$$

which coincides with the resistance term in eq. A.1 in ref. 7.

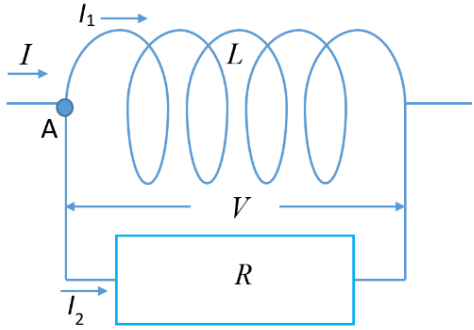

Fig. 15: Lumped circuit model representation of the current flow around a fluxon.

From the replacements (c.f. ref. 7 and Figs. 15 and 16)

$$R \to \frac{1}{\sigma_n \delta}; \quad L \to \mu_0 \lambda; \quad I \to \frac{B_{rf} \cdot w}{\mu_0} \quad ; Re(Z) \to R_{fl}$$

and, as $2 \cdot \sigma_n \cdot \omega \cdot \mu_0 \cdot \lambda^2 \ll 1$ as well as

$$\delta = \sqrt{\frac{2}{\sigma_n \cdot \omega \cdot \mu_0}}$$

one arrives at

$$R_{fl} = (\omega \mu_0)^{3/2} (2 \cdot \sigma_n)^{1/2} \lambda^2 \quad ,$$

showing clearly the dependence of $R_{fl}$ on frequency ($\sim \omega^{3/2}$).

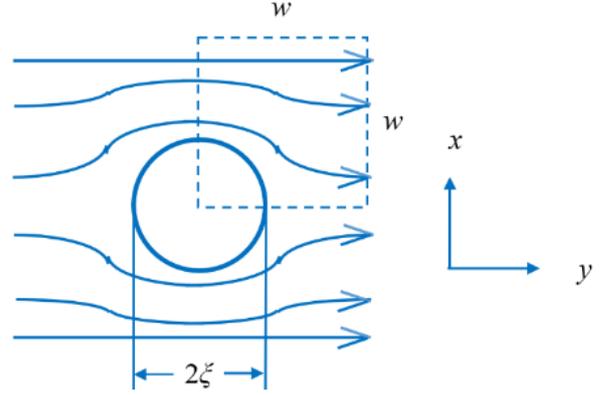

Fig. 16: Schematic current path around a fluxon: the current path is shown in plan view. The square of width $w$ shows a quarter of the perturbed region of current due to the presence of the fluxon of diameter $2 \cdot \xi$. The current penetrates the paper plane perpendicularly a distance $\delta$ into the fluxon, yet the current penetrates the sc metal in the vicinity of the fluxon to a distance of $\lambda$.

The total average power loss $P$ per cavity surface consists of the losses of an individual fluxon summed over the number $N$ of fluxons with individual area $w^2$,

$$P = \frac{1}{2} \cdot \sum_{i=1}^{N} w^2 R_{fl} \left(\frac{B_{rf}}{\mu_0}\right)^2 = \frac{1}{2} \cdot N w^2 R_{fl} \left(\frac{B_{rf}}{\mu_0}\right)^2.$$

As the magnetic flux $\Phi$ from the ambient magnetic field $B$ across the cavity area $A$, $\Phi = A \cdot B$, is (nearly) completely trapped upon cool down, the flux $\Phi$ is redistributed in the form of fluxons. Their number $N$ is defined by $\Phi = B_{c2} \cdot N \cdot w^2$. Hence the dissipated power $p$ per area $A$

$$p = \frac{P}{A} = \frac{1}{2} \cdot \underbrace{\frac{B}{B_{c2}} \cdot R_{fl}}_{R_s} \cdot \left(\frac{B_{rf}}{\mu_0}\right)^2 = \frac{1}{2} \cdot \underbrace{\frac{R_{fl}}{B_{c2}}}_{R_{fl}^0} \cdot B \cdot \left(\frac{B_{rf}}{\mu_0}\right)^2 \quad ,$$

defining the fluxon sensitivity $R_{fl}^0$ to

$$R_{fl}^0 = c \cdot (\omega \mu_0)^{3/2} (2 \cdot \sigma_n)^{1/2} \lambda^2 \frac{1}{B_{c2}} \quad . \quad \text{(A-2)}$$

By including the correction factor $c$ (62.5 %) for the cavity surface being only partially exposed to the perpendicular component of the external magnetic field eq. 4 is reproduced.

## $R_{fl}^0$ for N-doped niobium cavities

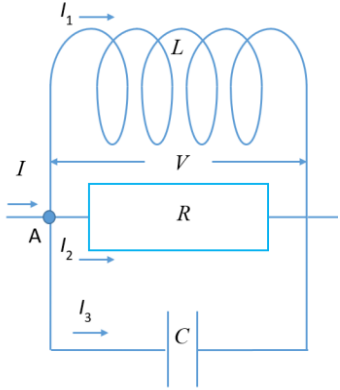

Fig. 17: Lumped circuit model representation of the current flow around a fluxon including a capacitance.

The supposition is made that in Fig. 15 a capacitance is added to account for possible polarisation charges in the N-doped niobium considered as a disordered composite [44] whose components have different electrical conductivities, Fif. 17. It is well known that such polarisation charges appear, by force of continuity of the RF current, in conductors with variable electrical conductivity or cross section. Therefore, eq. A-1 is complemented by a capacitance $C$,

$$\left(\frac{1}{i\omega L} + i\omega C + \frac{1}{R}\right) V(t) = I(t) \qquad . \qquad (A-3)$$

It is instructive to modify this equation as follows. The impedance is

$$Z = \frac{V}{I} = \frac{1}{\frac{1}{i\omega L} + i\omega C + \frac{1}{R}}$$

with the real part

$$Re(Z) = \frac{R}{\left[\left(\omega C - \frac{1}{\omega L}\right) \cdot R\right]^2 + 1} \qquad .$$

By inserting lumped circuit elements such as the resonant frequency $\omega_0 = 1/\sqrt{(L \cdot C)}$ and the factor of merit $Q = \omega_0 \cdot R \cdot C$ the real part of the impedance is rewritten as

$$Re(Z) = \frac{R}{\left[\left(\frac{\omega}{\omega_0} - \frac{\omega_0}{\omega}\right) \cdot Q\right]^2 + 1} \qquad . \qquad (A-4)$$

The damped oscillator equation A-3 has its equivalent in the equation of motion for the fluxon (eq. 2 in ref. 45),

$$\left(\frac{k}{i\omega} + i\omega M + \eta\right) \dot{x} = J_0 \cdot \Phi_0 \qquad . \qquad (A-5)$$

All these quantities are defined in Table 7.
By induction, a moving fluxon lattice of width w exposed to the trapped magnetic field $B$ is subject to the voltage

$$V = w\dot{x}B ,$$

such that eq. A-5 is modified to

$$\left(\frac{k}{i\omega} + i\omega M + \eta\right) V = J_0 \Phi_0 wB = I\Phi_0 B/\lambda \qquad (A-6)$$

with the total current $I = J_0 \cdot w \cdot \lambda$.

Table 7: Definition of relevant parameters.

| Name | Definition |
|---|---|
| Elastic constant per unit fluxon length (J/m³) | $k = \frac{2\pi}{d} \cdot \frac{\alpha \cdot \Phi_0}{B}$ |
| Mass per unit fluxon length (kg/m) | $M = 2 \cdot \pi \cdot n \cdot m \cdot \xi^2$ |
| Flow viscosity per unit length (Js/m³) | $\eta = \frac{\Phi_0 \cdot B_{c2}}{\rho}$ |
| Maximum Lorentz force the fluxon lattice can withstand (N/m³) | $\alpha = B \cdot J_c$ |
| Trapped magnetic field (Vs/m²) | $B = \frac{\Phi_0}{d^2}$ |
| $\dot{x}$: fluxon velocity (m/s) $d$: fluxon distance (m) $B_{c2}$: upper critical magnetic field (Vs/m²) $J_c$: critical current density (A/m²) $J_0$: RF current density (A/m²) $\rho$: electrical resistivity (Ωm) $n$: electron density (m⁻³) $m$: electron mass (kg) $\xi$: coherence length (m) $\Phi_0$: flux quantum (Vs) | |

Comparing eqs. A-3 and A-6 results in these definitions:

$$L = \frac{\Phi_0 \cdot B}{k \cdot \lambda} ; R = \frac{\Phi_0 \cdot B}{\eta \cdot \lambda} ; C = \frac{M \cdot \lambda}{\Phi_0 \cdot B} \qquad . \qquad (A-7)$$

The fluxon sensitivity is

$$R_{fl}^0 = c \cdot \frac{R}{1 + \left(\omega CR - \frac{R}{\omega L}\right)^2} \cdot \frac{1}{B_{c2}} \qquad . \qquad (A-8)$$

With the resonant frequency $\omega_0$, the factor of merit $Q$,

$$\omega_0 = \frac{1}{\sqrt{L \cdot C}} = \sqrt{\frac{k}{M}} \quad , \quad Q = \omega_0 RC = \frac{\sqrt{kM}}{\eta} \quad , (A-9)$$

and with $R$ as in eq. A-7, follows for the fluxon sensitivity

$$R_{fl}^0 = c \cdot \frac{R}{1 + \left[\left(\frac{\omega}{\omega_0} - \frac{\omega_0}{\omega}\right) \cdot Q\right]^2} \cdot \frac{1}{B_{c2}} \qquad . \qquad (A-10)$$

It should be noted that the resonant frequency $\omega_0$ and the depinning frequency $f_d$ (eq. 3) are defined differently. Eq. 3 can be transformed to

$$\omega_d = 2\pi f_d = \frac{k}{\eta} \qquad .$$

This same result could also be obtained from eq. A-8 through setting by supposition $M = 0$ (or equivalently $C = 0$) and replacing $R$ and $L$ as in eq. A-7:

$$\frac{R_{fl}^0(\omega_d)}{R_{fl}^0(\omega \to \infty)} = \frac{1}{2} = \frac{1}{1+\left(\frac{R}{\omega_d L}\right)^2} \quad,$$

$$\omega_d = \frac{R}{L} \to \frac{k}{\eta} \quad.$$

The following conclusion holds: for the mean free path interval under study and the numbers as listed in Table 6, the resonant frequencies $\omega_0$ are 5 - 40 times larger compared to the depinning frequencies $\omega_d$.

*Note added in proof*: As this section can easily be misunderstood, the following clarification may be helpful.

The common criterion for the anomalous skin effect is the very large mean free path length $l$ compared to the penetration depth $\lambda$. This is not the case in the present analysis. Nevertheless, the anomalous skin effect is postulated here in the sense that the electric field perpendicular to the surface (cf. $E_2$ in Fig. 5) is confined within a volume around the fluxon of lateral width of the order of the coherence length $\xi$. This means that $\xi$ and not $\lambda$ serves as criterion for the anomalous skin effect. Then the corresponding condition $l/\xi = 1+l/\xi_0 > 1$ is generally fulfilled ($\xi_0$ is the intrinsic coherence length).